\documentclass[twocolumn,showpacs,preprintnumbers,amsmath,amssymb,showkeys]{revtex4}

\usepackage{graphicx}
\usepackage{dcolumn}
\usepackage{bm}
\usepackage{braket}

\begin{document}

\noindent

\preprint{}

\title{Conservation of correlation in measurement underlying the violation of Bell inequalities and a game of joint mapping}

\author{Agung Budiyono}
\email{agungbymlati@gmail.com}
\affiliation{Department of Engineering Physics, Bandung Institute of Technology, Bandung, 40132, Indonesia} 
\affiliation{Research Center for Nanoscience and Nanotechnology, Bandung Institute of Technology, Bandung, 40132, Indonesia}

\date{\today}

\begin{abstract} 
What compels quantum measurement to violate the Bell inequalities? Suppose that regardless of measurement, one can assign to a spin-$\frac{1}{2}$ particle (qubit) a definite value of spin, called c-valued spin variable, but, it may take any continuous real number. Suppose further that measurement maps the c-valued spin variable from the continuous range of possible values onto the binary standard quantum spin values $\pm 1$ while preserving the bipartite correlation. Here, we show that such c-valued spin variables can indeed be constructed. In this model, one may therefore argue that it is the requirement of conservation of correlation which compels quantum measurement to violate the Bell inequalities when the prepared state is entangled. We then discuss a statistical game which captures the model of measurement, wherein two parties are asked to independently map a specific ensemble of pairs of real numbers onto pairs of binary numbers $\pm 1$, under the requirement that the correlation is preserved. The conservation of correlation forces the game to respect the Bell theorem, which implies that there is a class of games no classical (i.e., local and deterministic) strategy can ever win. On the other hand, a quantum strategy with an access to an ensemble of entangled spin-$\frac{1}{2}$ particles and circuits for local quantum spin measurement, can be used to win the game. 
\end{abstract} 

\pacs{03.65.Ta, 03.65.Ca}
\keywords{quantum measurement, entanglement, quantum correlation, violation of Bell inequalities, real c-valued spin variable, global-nonseparable random variable, conservation of correlation, game of joint mapping, nonlocality, indeterminism}
\maketitle       

\section{Introduction}

Bell showed in 1964 \cite{Bell's theorem} that certain correlations between the outcomes of two spacelike separated quantum measurements over entangled states violate his eponymous inequalities. Any model that reproduces the quantum correlation must therefore give up at least one of the plausible premises based on which the Bell inequalities are derived: predetermination or realism, locality, and free choice or no superdeterminism \cite{Bell's theorem,CHSH inequality}. See Ref. \cite{Wiseman Bell's two theorems} for other possible premises, and the different combination of the premises underlying the Bell inequalities. Hitherto, there is no general consensus as to the precise implication of the above Bell theorem concerning the nature of physical realities and/or the structure of causation underlying the microscopic phenomena \cite{Norsen nonlocality,Tumulka nonlocality,Gisin nonlocality,Hall nonlocality,Muynck CFI,Zukowski nonrealism,Zukowski CFI,Blaylock CFI,Fine-Bell theorem,Fine prism model,Fine prism model2,Jarrett incompleteness,Ballentine-Jarrett paper,Howard nonseparability,Fuchs Qbism,Griffiths consistent locality,Price backward causation,Spekkens fine tuning}. The alternative possible explanations of the violation of the Bell inequalities arguably cannot be separated from the different resolutions of the long standing measurement problem which is central in the debate about the meaning of quantum mechanics \cite{Bub book}. For example, Bohmian mechanics \cite{Bohmian mechanics}, which resolves the measurement problem by introducing a hidden variable determining the measurement outcomes, must have a gross nonlocality to comply with the Bell theorem. By contrast, Copenhagen interpretation rejects the presence of such variables, and argues that the violation of Bell inequalities does not imply nonlocality \cite{Peres instrumentalist credo,Peres instrumentalist book}.  

But, what compels measurement to violate the Bell inequalities? Are there some profound principles that measurement must obey so that it is willing to give up plausible and intuitive concepts such as locality \cite{Bohmian mechanics}, determinism \cite{Peres instrumentalist credo,Peres instrumentalist book}, or/and free choice \cite{Brans fully causal hidden variable model,Hall relaxing measurement independence}? In attempt to better understand this question, let us first suppose that, regardless of measurement, one can assign to a spin-$\frac{1}{2}$ particle (or, a generic qubit) a definite (i.e., determinate) value of spin, called c-valued spin variable. Moreover, let us assume that the c-valued spin variable may take on any continuous real number prior to the measurement, and the  spin measurement maps it onto the standard binary quantum spin values $\pm 1$. While measurement in general changes the c-valued spin variable, it is natural to require that measurement preserves the statistical correlation between the c-valued spin variables of two particles. Can real c-valued spin variables be constructed which meet the above conditions for measurement? A positive answer is given in the present work. 

In the above model of measurement, we may therefore argue that it is the requirement of conservation of correlation which compels the violation of Bell inequalities for entangled state with bizarre possible implications, in a similar fashion that assuming the constancy of speed of light for all inertial coordinate systems implies counterintuitive observable effects such as time dilation. To make the idea more transparent, we then construct a game of spacelike joint mappings as follows. Suppose that Alice and Bob, isolated from one another, are given a pair of real numbers, one pair at a time, sampled from a specific distribution associated with a vector in four dimensional complex Hilbert space. We ask them to independently map the pair of real numbers onto a pair of binary numbers $\pm 1$, with a constraint that their outputs must preserve the statistical correlation of the inputs. Is there a classical (i.e., local and deterministic) joint strategy for Alice and Bob to always win the game? We show that the requirement of conservation of correlation forces Alice and Bob's joint strategy to comply with the Bell theorem. This implies that for certain initial correlations associated with nonfactorizable vector in the Hilbert space, Alice and Bob will never win the game using any classical joint strategy, i.e., their mappings will violate the conservation of correlation. They can instead win the game by running a quantum strategy using an ensemble of entangled pair of spin-$\frac{1}{2}$ particles (qubits) and quantum circuits for local measurement of spin observables.        

\section{Real c-valued spin variables and conservation of bipartite correlation in measurement} 

In this article, for our purpose, we only consider systems of two spin-$\frac{1}{2}$ particles (or, a pair of arbitrary physical qubits), referred to as particle 1 and particle 2. First, we choose a set of basis vectors $\{\ket{\eta_{12}}\}$ of the four dimensional Hilbert space, so that $\sum_{\eta_{12}}\ket{\eta_{12}}\bra{\eta_{12}}=\hat{\mathbb{I}}_{12}$, where $\hat{\mathbb{I}}_{12}$ is the $4\times 4$ identity matrix. We refer to such a complete set of basis vectors as the reference coordinate basis, and assume that it is factorizable, i.e., $\{\ket{\eta_{12}}\}=\{\ket{\eta_1}\ket{\eta_2}\}$, where $\{\ket{\eta_{\mu}}\}$ is the complete basis of the Hilbert space associated with particle $\mu$, $\mu=1,2$. Then, regardless of any measurement, given a preparation represented by a pure quantum state $\ket{\psi_{12}}$ of the two particles, the `c-valued spin variable' associated with particle $\mu$ along a direction represented by a unit vector $\vec{n}_{\mu}$ in three dimensional space, within the reference basis $\{\ket{\eta_{12}}\}$ with $\braket{\eta_{12}|\psi_{12}}\neq 0$, is defined as follows:
\begin{eqnarray}
&&\tilde{s}_{\vec{n}_{\mu}}(\eta_{12},\xi|\psi_{12})\nonumber\\
&&\doteq{\rm Re}\Big\{\frac{\braket{\eta_{12}|\hat{\sigma}_{\vec{n}_{\mu}}|\psi_{12}}}{\braket{\eta_{12}|\psi_{12}}}\Big\}+\frac{\xi}{\hbar}{\rm Im}\Big\{\frac{\braket{\eta_{12}|\hat{\sigma}_{\vec{n}_{\mu}}|\psi_{12}}}{\braket{\eta_{12}|\psi_{12}}}\Big\}. 
\label{real continuum c-valued spin variable} 
\end{eqnarray}
Here, $\hat{\sigma}_{\vec{n}_{\mu}}\doteq\vec{n}_{\mu}\cdot\vec{\hat{\sigma}}$, where $\vec{\hat{\sigma}}=(\hat{\sigma}_x,\hat{\sigma}_y,\hat{\sigma}_z)$ is the vector of the Pauli operators, and $\xi$ is a real-valued global-nonseparable variable. $\hat{\sigma}_{\vec{n}_{\mu}}$ in the numerator is the short hand for $\hat{\sigma}_{\vec{n}_{\mu}}\otimes\hat{\mathbb{I}}_{\nu}$, $\mu\neq\nu$, $\mu,\nu=1,2$, where $\hat{\mathbb{I}}_{\nu}$ is the $2\times 2$ identity matrix of the Hilbert space of the particle $\nu$.  

We further assume that the probability distribution for the coordinate value $\eta_{12}$ follows the Born's rule, i.e., 
\begin{eqnarray}
{\rm Pr}(\eta_{12}|\psi_{12})=|\braket{\eta_{12}|\psi_{12}}|^2.
\label{Born's rule}
\end{eqnarray} 
Moreover, $\xi$ is assumed to fluctuate randomly on a microscopic time scale independent of the prepared quantum state $\ket{\psi_{12}}$, the spin observable $\hat{\sigma}_{\vec{n}_{\mu}}$ and the reference basis $\{\ket{\eta_{12}}\}$, with its first two moments are given by 
\begin{eqnarray}
\overline{\xi}\doteq\sum_{\xi}\xi\chi(\xi)=0,~~ \overline{\xi^2}=\hbar^2, 
\label{Planck constant}
\end{eqnarray}
where $\chi(\xi)$ is the probability distribution of $\xi$, and the summation is replaced by a suitable integration for a continuous $\xi$. The ensemble average of any function $f(\tilde{s}_{\vec{n}_1},\tilde{s}_{\vec{n}_2})$ of two c-valued spin variables for a given prepraration $\ket{\psi_{12}}$ is then defined as in the conventional probability theory: 
\begin{eqnarray}
\braket{f(\tilde{s}_{\vec{n}_1}(\eta_{12},\xi|\psi_{12}),\tilde{s}_{\vec{n}_2}(\eta_{12},\xi|\psi_{12}))}\nonumber\\
\doteq\sum_{\eta_{12}}\sum_{\xi}f(\tilde{s}_{\vec{n}_1},\tilde{s}_{\vec{n}_2})\chi(\xi){\rm Pr}(\eta_{12}|\psi_{12}). 
\label{ensemble average}
\end{eqnarray}

The above definition of c-valued spin variables can be extended to general quantum observables acting on general quantum states \cite{Agung general c-valued physical quantities and uncertainty relation}. It was initially conceived for phase space variables to study a specific epistemic (i.e., statistical) restriction underlying the incompatibility between quantum observables for position and momentum \cite{Agung ERPS representation,Agung-Daniel model}. We note that the two terms on the right-hand side of Eq. (\ref{real continuum c-valued spin variable}) can be operationally interpreted respectively as the real and imaginary part of weak value obtained in weak measurement with postselectdion \cite {Aharonov weak value,Aharonov-Daniel book,Lundeen complex weak value,Jozsa complex weak value}. They can also be interpreted respectively as the optimal estimate of the left-hand side and the associated estimation error \cite{Agung epistemic interpretation,Agung estimation independence,Hall exact UR,Johansen weak value best estimation,Hall prior information,Hofmann imaginary part of weak value in optimal estimation}. However, in this work, we are not concerned with such operational interpretations. Rather, the c-valued spin variable is defined independently of any kind of measurement. Hence, unlike the weak value and the scheme of optimal estimation above, in this article, the reference basis is fixed once and for all, i.e., it cannot be varied freely by the experimenter.   

Notice first that, unlike the standard quantum spin values, given $\ket{\psi_{12}}$, $\{\ket{\eta_{12}}\}$ and $\xi$, the c-valued spin variable $\tilde{s}_{\vec{n}_{\mu}}(\eta_{12},\xi|\psi_{12})$ defined in Eq. (\ref{real continuum c-valued spin variable}) for any direction $\vec{n}_{\mu}$ has always definite value. Moreover, $\tilde{s}_{\vec{n}_{\mu}}(\eta_{12},\xi|\psi_{12})$ may take on continuum real numbers depending on the continuous parameterization of $\ket{\psi_{12}}$ and $\vec{n}_{\mu}$ associated with the spin observable $\hat{\sigma}_{\vec{n}_{\mu}}$, as expected for variables in classical mechanics; see an example below. Next, the value assignment of $\tilde{s}_{\vec{n}_{\mu}}(\eta_{12},\xi|\psi_{12})$ of particle $\mu$ depends in general on the global prepared quantum state $\ket{\psi_{12}}$ of the two particles. In the specific case when the prepared quantum state is factorizable, i.e., $\ket{\psi_{12}}=\ket{\psi_1}\ket{\psi_2}$, where $\ket{\psi_1}$ and $\ket{\psi_2}$ are the quantum states associated with the independent preparation of the particle 1 and particle 2, respectively, then, noting that $\frac{\braket{\eta_{12}|\hat{\sigma}_{\vec{n}_{\mu}}|\psi_{12}}}{\braket{\eta_{12}|\psi_{12}}}=\frac{\braket{\eta_{\mu}|\hat{\sigma}_{\vec{n}_{\mu}}|\psi_{\mu}}}{\braket{\eta_{\mu}|\psi_{\mu}}}$, $\mu=1,2$, one has  
\begin{eqnarray}
&&\tilde{s}_{\vec{n}_{\mu}}(\eta_{12},\xi|\psi_{12})\nonumber\\
&=&{\rm Re}\Big\{\frac{\braket{\eta_{\mu}|\hat{\sigma}_{\vec{n}_{\mu}}|\psi_{\mu}}}{\braket{\eta_{\mu}|\psi_{\mu}}}\Big\}+\frac{\xi}{\hbar}{\rm Im}\Big\{\frac{\braket{\eta_{\mu}|\hat{\sigma}_{\vec{n}_{\mu}}|\psi_{\mu}}}{\braket{\eta_{\mu}|\psi_{\mu}}}\Big\}\nonumber\\
&=&\tilde{s}_{\vec{n}_{\mu}}(\eta_{\mu},\xi|\psi_{\mu}), 
\label{real-global local deterministic continuum c-valued spin for separable state} 
\end{eqnarray}
namely, given the value of $\xi$, the c-valued spin variable associated with particle $\mu$ is independent of that associated with particle $\nu$, $\mu\neq\nu$. However, even when the two particles are independently prepared, the c-valued spins associated with the two particles, i.e., $\tilde{s}_{\vec{n}_1}(\eta_1,\xi|\psi_1)$ and $\tilde{s}_{\vec{n}_2}(\eta_2,\xi|\psi_2)$, are in general instantaneously connected via the global variable $\xi$ regardless of their spatial distance. There is an exception. When $\ket{\psi_{\mu}}$ in Eq. (\ref{real-global local deterministic continuum c-valued spin for separable state}) is given by one of the eigenstates of $\hat{\sigma}_{\vec{n}_{\mu}}$, which is just the case after the measurement of $\hat{\sigma}_{\vec{n}_{\mu}}$, the second term on the right-hand side vanishes. Moreover, the first term is exactly equal to the eigenvalue $o_{\vec{n}_{\mu}}$ of $\hat{\sigma}_{\vec{n}_{\mu}}$ so that we have $\tilde{s}_{\vec{n}_{\mu}}=o_{\vec{n}_{\mu}}=\pm 1$ independent of $\xi$. Hence, in this specific case, the two c-valued spins associated with two independently prepared particles, are fully independent of each other, given by the standard quantum spin values.   

Next, despite the c-valued spin variables are always determinate in the absence of measurement, they satisfy a complementarity principle, in the following sense. Consider two spin operators $\hat{\sigma}_{\vec{n}_{\mu}}$ and $\hat{\sigma}_{\vec{n}'_{\mu}}$ associated with particle $\mu$, so that $[\hat{\sigma}_{\vec{n}_{\mu}},\hat{\sigma}_{\vec{n}'_{\mu}}]\neq 0$, $\mu=1,2$. Then, for any preparation $\ket{\psi_{12}}$, the associated c-valued spin variables, i.e., $\tilde{s}_{\vec{n}_{\mu}}(\eta_{12},\xi|\psi_{12})$ and $\tilde{s}_{\vec{n}'_{\mu}}(\eta_{12},\xi|\psi_{12})$, cannot simultaneously equal to $\pm 1$ independent of $\xi$ \cite{Agung general c-valued physical quantities and uncertainty relation}. For example, if  for a given $\ket{\psi_{12}}$ we have $\tilde{s}_{\vec{n}_{\mu}}(\eta_{12},\xi|\psi_{12})=\pm 1$ independent of $\xi$ which is the case when $\ket{\psi_{12}}$ is the eigenstate of $\hat{\sigma}_{\vec{n}_{\mu}}$, then we must have $\tilde{s}_{\vec{n}'_{\mu}}(\eta_{12},\xi|\psi_{12})\neq\pm 1$ fluctuating randomly with $\xi$, and vice versa. This captures the quantum complementarity between $\hat{\sigma}_{\vec{n}_{\mu}}$ and $\hat{\sigma}_{\vec{n}'_{\mu}}$ in the Copenhagen interpretation, that is, the two noncommuting spin operators cannot be jointly measured, thus assigned $\pm 1$ values, simultaneously. Indeed, like the standard quantum spin values, the variances of the c-valued spin variables defined in Eq. (\ref{real continuum c-valued spin variable}) satisfy the Heisenberg-Kennard-Robertson-Schr\"odinger uncertainty relation \cite{Agung general c-valued physical quantities and uncertainty relation}. 

The above observation shows that the c-valued spin variables defined in Eq. (\ref{real continuum c-valued spin variable}) share many qualitative features of the standard quantum spin values. Moreover, while the exact value of the c-valued spin variable depends on the choice of reference basis, its qualitative features do not. This suggests that the c-valued spin variables can be seen as a natural extension of the standard quantum spin values to the situation when there is no measurement. Additionally, as shown below, regardless of the choice of the reference basis, the local average and bipartite correlation of the c-valued spin variables prior to measurement are equal respectively to the local average and bipartite correlation of the associated standard quantum spin values obtained in measurement. 

For illustration and later reference, let us first consider the case when the prepared quantum state of the pair of the particles is given by the singlet, i.e.,   
\begin{eqnarray}
\ket{\psi_{12}^{\mathcal S}}\doteq\frac{1}{\sqrt{2}}(\ket{01}-\ket{10}),
\label{singlet state}
\end{eqnarray} 
where $\{\ket{0},\ket{1}\}$ are the eigenstates of $\hat{\sigma}_{z}$. Let us choose the following complete set of vectors as the reference basis: $\{\ket{y+}\ket{x+},\ket{y+}\ket{x-},\ket{y-}\ket{x+},\ket{y-}\ket{x-}\}$, where $\ket{x\pm}=\frac{1}{\sqrt{2}}(\ket{0}\pm\ket{1})$ and $\ket{y\pm}=\frac{1}{\sqrt{2}}(\ket{0}\pm i\ket{1})$. Moreover, without loosing generality, assume that the spin operator of the first particle lies on the $xz$-plane with a polar angle $\theta_1$ with respect to the positive $z$-axis. Computing the c-valued spin $\tilde{s}_{\vec{n}_{\theta_1}}$ defined in Eq. (\ref{real continuum c-valued spin variable}), one obtains 
\begin{eqnarray}
\tilde{s}_{\vec{n}_{\theta_1}}(y+,x+,\xi|\psi_{12}^{\mathcal S})&=&-\sin\theta_1-\frac{\xi}{\hbar}\cos\theta_1;\nonumber\\
\tilde{s}_{\vec{n}_{\theta_1}}(y+,x-,\xi|\psi_{12}^{\mathcal S})&=&\sin\theta_1+\frac{\xi}{\hbar}\cos\theta_1;\nonumber\\
\tilde{s}_{\vec{n}_{\theta_1}}(y-,x+,\xi|\psi_{12}^{\mathcal S})&=&-\sin\theta_1+\frac{\xi}{\hbar}\cos\theta_1;\nonumber\\
\tilde{s}_{\vec{n}_{\theta_1}}(y-,x-,\xi|\psi_{12}^{\mathcal S})&=&\sin\theta_1-\frac{\xi}{\hbar}\cos\theta_1. 
\label{c-valued spin for along theta for singlet}
\end{eqnarray}
Hence, it varies continuously with the direction of the spin observable parameterized by $\theta_1$, as classical angular momentum. Let us proceed to consider the case when $\vec{n}_1=\vec{n}=\vec{n}_2$, i.e., the spin observables of the two particles are pointing along the same direction. Then, noting that $(\hat{\sigma}_{\vec{n}_1}\otimes\hat{\mathbb{I}}_2)\ket{\psi_{12}^{\mathcal S}}=-(\hat{\mathbb{I}}_1\otimes\hat{\sigma}_{\vec{n}_2})\ket{\psi_{12}^{\mathcal S}}$, and inserting into Eq. (\ref{real continuum c-valued spin variable}), we have 
\begin{eqnarray}
\tilde{s}_{\vec{n}_1}(\eta_{12},\xi|\psi^{\mathcal S}_{12})=-\tilde{s}_{\vec{n}_2}(\eta_{12},\xi|\psi^{\mathcal S}_{12}),
\label{opposite spinning pair for an singlet state}
\end{eqnarray}
i.e., they are always perfectly anti-correlated like the associated standard quantum spin values. Hence, the conservation of spin angular momentum for singlet holds even in the absence of measurement, as expected in classical mechanics. 

Now, consider the case when $\vec{n}_1$ and $\vec{n}_2$ are coplanar lying on the $xz$-plane tilted from the positive $z$-axis with polar angles respectively given by $\theta_1$ and $\theta_2$. Computing the correlation between the c-valued spins $\tilde{s}_{\vec{n}_{\theta_1}}$ and $\tilde{s}_{\vec{n}_{\theta_2}}$ associated with the singlet state, one recovers, using Eqs. (\ref{c-valued spin for along theta for singlet}) and noting (\ref{opposite spinning pair for an singlet state}), the correlation between the associated standard quantum spin values for the singlet state which violates the Bell inequalities:  
\begin{eqnarray}
&&\braket{\tilde{s}_{\vec{n}_{\theta_1}}(\eta_{12},\xi|\psi^{\mathcal{S}}_{12})\tilde{s}_{\vec{n}_{\theta_2}}(\eta_{12},\xi|\psi^{\mathcal{S}}_{12})}\nonumber\\
&&\doteq\sum_{\eta_{12}}\sum_{\xi}~\tilde{s}_{\vec{n}_1}(\eta_{12},\xi|\psi^{\mathcal{S}}_{12})\tilde{s}_{\vec{n}_2}(\eta_{12},\xi|\psi^{\mathcal{S}}_{12})\nonumber\\
&&\hspace{5mm}\times\chi(\xi){\rm Pr}(\eta_{12}|\psi_{12})\nonumber\\
&&=-\sin\theta_1\sin\theta_2-\cos\theta_1\cos\theta_2=-\cos(\theta_2-\theta_1)\nonumber\\
&&=\braket{\psi^{\mathcal{S}}_{12}|(\hat{\sigma}_{\vec{n}_{\theta_1}}\otimes\hat{\sigma}_{\vec{n}_{\theta_2}})|\psi^{\mathcal{S}}_{12}},  
\end{eqnarray}
where we have used Eq. (\ref{Planck constant}) and noting that ${\rm Pr}(\pm y,\pm x|\psi^{\mathcal{S}}_{12})=|\braket{\pm y,\pm x|\psi^{\mathcal{S}}_{12}}|^2=1/4$. 

Indeed, the above result applies to general cases as stated by the following theorem. \\
{\bf Theorem 1}:\\
The statistical correlation between two continuum c-valued spins $\tilde{s}_{\vec{n}_1}$ and $\tilde{s}_{\vec{n}_2}$ along any pair of directions $(\vec{n}_1,\vec{n}_2)$ within any reference basis $\{\ket{\eta_{12}}\}$ for arbitrary prepared quantum state $\ket{\psi_{12}}$, is equal to the correlation between the binary standard quantum spin values obtained from the measurement of the quantum spin observables $(\hat{\sigma}_{\vec{n}_1}\otimes\hat{\sigma}_{\vec{n}_2})$ over $\ket{\psi_{12}}$, i.e.,   
\begin{eqnarray}
\braket{\tilde{s}_{\vec{n}_1}(\eta_{12},\xi|\psi_{12})\tilde{s}_{\vec{n}_2}(\eta_{12},\xi|\psi_{12})}=\braket{\psi_{12}|\hat{\sigma}_{\vec{n}_1}\otimes\hat{\sigma}_{\vec{n}_2}|\psi_{12}}. 
\label{Theorem 1}
\end{eqnarray}
This theorem is a special case of a theorem presented in the previous work \cite{Agung general c-valued physical quantities and uncertainty relation}. Moreover, taking $\hat{\sigma}_{\nu}=\hat{\mathbb{I}}_{\nu}$, and noting that $\tilde{\mathbb{I}}_{\nu}(\eta_{12},\xi|\psi_{12})=1$, we have $\braket{\tilde{s}_{\vec{n}_\mu}(\eta_{12},\xi|\psi_{12})}=\braket{\psi_{12}|\hat{\sigma}_{\vec{n}_\mu}\otimes\hat{\mathbb{I}}_{\nu}|\psi_{12}}={\rm Tr}_{\mu}\{\hat{\sigma}_{\vec{n}_\mu}\hat{\varrho}_{\mu}\}$, where $\hat{\varrho}_{\mu}={\rm Tr}_{\nu}\{\ket{\psi_{12}}\bra{\psi_{12}}\}$, $\nu\neq\mu$, i.e., the local ensemble average of the c-valued spin variable for any $\ket{\psi_{12}}$ is also equal to the local quantum expectation value.      

We note that, crucially, to arrive at the equality of Eq. (\ref{Theorem 1}), $\xi$ must be indeed global-nonseparable. Such a global-nonseparable variable $\xi$ presumes a preferred spacetime reference frame violating the Lorentz invariance underlying the theory of relativity. Next, at first sight, due to the nonseparability of $\xi$ which connects instantaneously the two c-valued spins, the model apparently will not be able to reconstruct the quantum correlation when the two particles are independently prepared so that the associated quantum state is factorizable, i.e. $\ket{\psi_{12}}=\ket{\psi_1}\ket{\psi_2}$. Remarkably, however, this is not the case since Theorem 1 applies for general quantum states, factorizable or not. Finally, note that Eq. (\ref{Theorem 1}) still applies if we replace the c-valued spin variables on the left hand side with the associated weak value of spin \cite{Hosoya-Shikano counterfactual value,Hall weak value to quantum uncertainty}. However, the weak values may take complex values when $\ket{\psi_{12}}$ is entangled (one can show that when $\ket{\psi_{12}}$ is factorizable, the real part of the weak values are sufficient to reconstruct the quantum correlation). In contrast to this, the c-valued spin variables defined in Eq. (\ref{real continuum c-valued spin variable}), which are always real, allow for the reconstruction of the quantum spins correlation for arbitrary quantum states at the cost of introducing the global-nonseparable variable $\xi$.  

Now, according to the standard quantum mechanics, the measurement of $\hat{\sigma}_{\vec{n}_{\mu}}$ inevitably projects the prepared quantum state $\ket{\psi_{12}}$ randomly onto one of the eigenstates $\ket{\phi_{12}^{\vec{n}_{\mu}}}$ of $\hat{\sigma}_{\vec{n}_{\mu}}$, i.e., $\ket{\psi_{12}}\mapsto \ket{\phi_{12}^{\vec{n}_{\mu}}}$, with the measurement outcomes given by the associated eigenvalues $o_{\vec{n}_{\mu}}=\pm 1$. Moreover, recall that evaluating the associated c-valued spin variable defined in Eq. (\ref{real continuum c-valued spin variable}) for these post-measurement quantum states $\ket{\phi_{12}^{\vec{n}_{\mu}}}$, we regain the standard quantum spin values, i.e., $\tilde{s}_{\vec{n}_{\mu}}(\eta_{12},\xi|\phi_{12}^{\vec{n}_{\mu}})=o_{\vec{n}_{\mu}}=\pm 1$. This observation suggests a model for the quantum spin measurement wherein it maps the c-valued spin variables from the continuous range of possible real values prior to measurement, onto the binary values $\pm 1$, i.e., 
\begin{eqnarray}
\mathbb{R}\ni\tilde{s}_{\vec{n}_{\mu}}(\eta_{12},\xi|\psi_{12})\mapsto\tilde{s}_{\vec{n}_{\mu}}(\eta_{12},\xi|\phi_{12}^{\vec{n}_{\mu}})\in\{-1,1\}, 
\label{mapping from continuum real number to binary in quantum measurement}
\end{eqnarray}
while preserving the bipartite correlation as per Theorem 1. Hence, we have upgraded the conservation of bipartite correlation as a principle which governs the measurement. In such a model, the restriction imposed by the violation of Bell inequalities to the statistics of the measurement outcomes $\pm 1$ for entangled state, thus arises from, and compelled by, the requirement of conservation of correlation in measurement.    

A few remarks are in order. First, let us emphasize that it is the statistical correlation that is preserved by the measurement, not the value assignment of the c-valued spin variables. We have thus relaxed the requirement for measurement in classical mechanics: i.e., rather than revealing the values of the variables prior to measurement, it is only required to reveal the bipartite correlation (and also the local average) prior to measurement. Hence, the measurement induced disturbance must comply with the principle of conservation of bipartite correlation. Next, when the prepared state is entangled, one finds that there is a nonlocal dependence of the value assignment of the c-valued spin variable of one particle on the spin measurement of the other (possibly arbitrarily remote) particle. For example, when the prepared state is a singlet, a spin measurement along the direction $\vec{n}_2=\vec{n}$ of particle $2$ with the outcome $+1(-1)$, will need to be accompanied by the mapping of $\tilde{s}_{\vec{n}_1}(\eta_{12},\xi|\psi_{12}^{\mathcal{S}})$ assigned to the particle $1$, where $\vec{n}_1=\vec{n}$, from its value prior to measurement given by Eq. (\ref{c-valued spin for along theta for singlet}), onto binary standard quantum spin values $-1(+1)$. However, since the statistics of the standard quantum spin values follows the Born's rule, such a nonlocal value assignment cannot be used for signalling. 

Hence, we have assumed that quantum bipartite correlation exists prior to measurement in terms of the correlation between the real c-valued spin variables. Remarkably, the c-valued spin variables can be constructed operationally via weak measurement with postselection and a classical postprocessing involving $\xi$. This correlation between the real c-valued spin variables already cannot be explained locally in terms of the correlation between classical variables in spacetime due to the dependence of the c-valued spin variables on the fluctuations of the global variable $\xi$. Joint spin measurement of the two particles preserves the correlation. Moreover, after the measurement, the dependence of the c-valued spin variables (now equal to the standard quantum spin values) on the global variable $\xi$ disappears. But, the nonclassicality reappears in the form of a nonlocal dependence of c-valued spin variable of one particle on the measurement of the other remote particle. 

\section{A game of joint mapping under conservation of correlation}

Bell theorem is most eloquently described in terms of spacelike coordination games which smartly exploit the classically counterintuitive features of quantum entanglement. For example, in the well-known CHSH game \cite{CHSH inequality,Jennings-Leifer review paper - nonclassicality}, Alice and Bob, spatially separated from each other, are required to independently come out with a pair of outputs based on a pair of inputs given randomly by a referee, Charlie, so that the outputs and inputs satisfy a simple arithmatic relation: 
\begin{eqnarray}
xy=a+b~ ({\rm mod}~2). 
\label{CHSH condition}
\end{eqnarray}
Here, $a$ is Alice's output given input $x$, and $b$ is Bob's output given input $y$, where $(x,y,a,b)=\{0,1\}$. Namely, to win the game, Alice and Bob must pop out different outputs, i.e., $a\neq b$, when their inputs are $x=y=1$, and pop out the same output, i.e., $a=b$, when at least one of their inputs is 0. One can show that if they only use classical (i.e., local and deterministic) joint strategy, their winning probability is lower than or equal to $3/4$ (assuming that all the four combinations of the inputs are equally sampled), a form of Bell inequalities. Surprisingly, if Alice and Bob share an ensemble of entangled qubits and have access to local spin measurement devices, they can win the game with a larger probability, as large as $(2+\sqrt{2})/4$ \cite{Tsirelson bound}. 

While the above game and other similar games \cite{Aravind game} strikingly show that entanglement is a nonclassical resource in certain protocols of information processing involving spacelike separated parties, the apparently mathematically simple winning condition of Eq. (\ref{CHSH condition}) is difficult to fathom in physical terms. What does the condition of Eq. (\ref{CHSH condition}) tell us about Nature so that it distinguishes quantum strategy from the classical strategy? Can we develop a different game with a winning condition that forces the violation of the Bell inequalities, which is more transparent and physically plausible, so that it can be upgraded as an axiom? Moreover, in the CHSH game, quantum measurement is treated as a total black box \cite{Popescu - Daniel PR box,Popescu review on PR box}, so that the physical constraint which compels the measurement to violate the Bell inequalities is not clear. Note that, within this point of view, nonlocality or/and indeterminism are not the constraints which force the measurement to violate the Bell inequalities, rather, they are the tricks that are possibly used by the measurement to satisfy the constraint. They (like time dilation in the theory of special relativity) should not therefore be upgraded as axioms, rather they are the implications of a deeper physical constraint. But, what is this deep physical constraint?  

Here, with Theorem 1 in mind, we construct a different two parties coordination game which to an extent captures the model of measurement speculated in the last paragraphs of the previous Section, as follows. First, a referee, Charlie, and two players, Alice and Bob, situated sufficiently faraway from each other, agree on a choice of a complex valued vector $\ket{\psi_{12}}$ in the computational basis. At each round of the game, Charlie samples a pair of random variables $(\eta_{12},\xi)$ from the joint probability distribution ${\rm Pr}(\eta_{12},\xi|\chi,\psi_{12})=|\braket{\eta_{12}|\psi_{12}}|^2\chi(\xi)$. Charlie then randomly chooses a pair of unit vectors, denoted respectively by $\vec{n}_1$ and $\vec{n}_2$, and use them, to compute $\tilde{s}_{\vec{n}_1}(\eta_{12},\xi|\psi_{12})$ and $\tilde{s}_{\vec{n}_2}(\eta_{12},\xi|\psi_{12})$ using the prescription in Eq. (\ref{real continuum c-valued spin variable}). In this way, these sets of numbers are effectively sampled from the joint probability distribution  
\begin{eqnarray}
&&{\rm Pr}(\tilde{s}_{\vec{n}_1},\tilde{s}_{\vec{n}_2},\eta_{12},\xi|\vec{n}_1,\vec{n}_2,\psi_{12})\nonumber\\
&&=\delta\Big(\tilde{s}_{\vec{n}_1};{\rm Re}\Big\{\frac{\braket{\eta_{12}|\hat{\sigma}_{\vec{n}_1}|\psi_{12}}}{\braket{\eta_{12}|\psi_{12}}}\Big\}+\frac{\xi}{\hbar}{\rm Im}\Big\{\frac{\braket{\eta_{12}|\hat{\sigma}_{\vec{n}_1}|\psi_{12}}}{\braket{\eta_{12}|\psi_{12}}}\Big\}\Big)\nonumber\\
&&\times\delta\Big(\tilde{s}_{\vec{n}_2};{\rm Re}\Big\{\frac{\braket{\eta_{12}|\hat{\sigma}_{\vec{n}_2}|\psi_{12}}}{\braket{\eta_{12}|\psi_{12}}}\Big\}+\frac{\xi}{\hbar}{\rm Im}\Big\{\frac{\braket{\eta_{12}|\hat{\sigma}_{\vec{n}_2}|\psi_{12}}}{\braket{\eta_{12}|\psi_{12}}}\Big\}\Big)\nonumber\\
&&\times\chi(\xi)\big|\braket{\eta_{12}|\psi_{12}}\big|^2. 
\label{join factorizable distribution of continuum c-valued spins}
\end{eqnarray} 
Here $\delta(k;l)$ is the Kroneker delta, i.e., $\delta(k;l)=1$ if $k=l$, and $\delta(k;l)=0$ if $k\neq l$. 

Next, Charlie sends the triple of random variables $\{\eta_{12},\xi,\tilde{s}_{\vec{n}_1}\}$ to Alice, and $\{\eta_{12},\xi,\tilde{s}_{\vec{n}_2}\}$ to Bob. Given all the above information, Alice and Bob joint task is to pick up a pair of binary numbers, either $1$ or $-1$, independently of each other. Hence, denoting Alice's output as a binary random variable $o_{\vec{n}_1}$ and that of Bob's as $o_{\vec{n}_2}$, their task is essentially to independently map the pair of real numbers $(\tilde{s}_{\vec{n}_1},\tilde{s}_{\vec{n}_2})$ onto a pair of binary numbers $(o_{\vec{n}_1},o_{\vec{n}_2})$, i.e.,  
\begin{eqnarray}
\mathcal{F}_{12}[\psi_{12},\eta_{12},\xi]:(\tilde{s}_{\vec{n}_1},\tilde{s}_{\vec{n}_2})\mapsto (o_{\vec{n}_1},o_{\vec{n}_2}), 
\label{coordinate mapping}
\end{eqnarray}
where $\tilde{s}_{\vec{n}_{\mu}}\in\mathbb{R}$ and $o_{\vec{n}_{\mu}}\in\{+1,-1\}$, $\mu=1,2$, and $\mathcal{F}_{12}$ describes their joint strategy. They can devise any classical algorithm or strategy to accomplish the task and program it to their computational devices together before they are moving separately to their laboratories. Since Alice and Bob do the mappings independently of each other, the conditional probability that Alice pops out $o_{\vec{n}_1}$ is statistically independent of $o_{\vec{n}_2}$ and $\tilde{s}_{\vec{n}_2}$, i.e., ${\rm Pr}(o_{\vec{n}_1}|o_{\vec{n}_2},\tilde{s}_{\vec{n}_1},\tilde{s}_{\vec{n}_2},\eta_{12},\xi,\psi_{12})={\rm Pr}(o_{\vec{n}_1}|\tilde{s}_{\vec{n}_1},\eta_{12},\xi,\psi_{12})$, and likewise that Bob pops out $o_{\vec{n}_2}$ is independent of $o_{\vec{n}_1}$ and $\tilde{s}_{\vec{n}_1}$, i.e., ${\rm Pr}(o_{\vec{n}_2}|o_{\vec{n}_1},\tilde{s}_{\vec{n}_1},\tilde{s}_{\vec{n}_2},\eta_{12},\xi,\psi_{12})={\rm Pr}(o_{\vec{n}_2}|\tilde{s}_{\vec{n}_2},\eta_{12},\xi,\psi_{12})$, so that we have the factorizability condition:
\begin{eqnarray}
&&{\rm Pr}(o_{\vec{n}_1},o_{\vec{n}_2}|\tilde{s}_{\vec{n}_1},\tilde{s}_{\vec{n}_2},\eta_{12},\xi,\psi_{12})\nonumber\\
&=&{\rm Pr}(o_{\vec{n}_1}|\tilde{s}_{\vec{n}_1},\eta_{12},\xi,\psi_{12}){\rm Pr}(o_{\vec{n}_2}|\tilde{s}_{\vec{n}_2},\eta_{12},\xi,\psi_{12}). 
\label{factorizability condition}
\end{eqnarray}  

We then say they win the game if the statistical correlation between $o_{\vec{n}_1}$ and $o_{\vec{n}_2}$, obtained by repeating the above protocol (in principle infinitely) many times, are equal to the initial correlation between $\tilde{s}_{\vec{n}_1}$ and $\tilde{s}_{\vec{n}_2}$, i.e., 
\begin{eqnarray}
&&\sum_{(\eta_{12},\xi)}\sum_{(\tilde{s}_{\vec{n}_1},\tilde{s}_{\vec{n}_2})}\sum_{(o_{\vec{n}_1},o_{\vec{n}_2})}~o_{\vec{n}_1}o_{\vec{n}_2}\nonumber\\
&&\times{\rm Pr}(o_{\vec{n}_1},o_{\vec{n}_2}|\tilde{s}_{\vec{n}_1},\tilde{s}_{\vec{n}_2},\eta_{12},\xi,\psi_{12})\nonumber\\
&&\times{\rm Pr}(\tilde{s}_{\vec{n}_1},\tilde{s}_{\vec{n}_2},\eta_{12},\xi|\vec{n}_1,\vec{n}_2,\psi_{12})\nonumber\\
&&=\braket{\tilde{s}_{\vec{n}_1}(\eta_{12},\xi|\psi_{12})\tilde{s}_{\vec{n}_2}(\eta_{12},\xi|\psi_{12})}. 
\label{conservation of correlation}
\end{eqnarray}
To summarize, what Alice and Bob have to do is to independently map the pair of random variables $(\tilde{s}_{\vec{n}_1},\tilde{s}_{\vec{n}_2})$ which may take any continuous real numbers depending on the choice of $(\vec{n}_1,\vec{n}_2)$, onto binary random variables $(o_{\vec{n}_1},o_{\vec{n}_2})$, based on a joint strategy, so that the resulting correlation between $o_{\vec{n}_1}$ and $o_{\vec{n}_2}$ preserves the initial correlation between $\tilde{s}_{\vec{n}_1}$ and $\tilde{s}_{\vec{n}_2}$. 

We argue below that there is a class of games wherein no classical joint strategy can ever win as stated by the following theorem.\\
{\bf Theorem 2}:\\
For a class of games with initial value of correlation $\braket{\tilde{s}_{\vec{n}_1}(\eta_{12},\xi|\psi_{12})\tilde{s}_{\vec{n}_2}(\eta_{12},\xi|\psi_{12})}$ associated with a nonfactorizable complex vector $\ket{\psi_{12}}$, no classical (i.e., local and deterministic) joint strategy of Alice and Bob will win the spacelike game of joint mapping, i.e. their mappings must violate the conservation of correlation of Eq. (\ref{conservation of correlation}). \\
{\bf Proof:}\\
First, combining  Eq. (\ref{conservation of correlation}) with Eq. (\ref{Theorem 1}) of the Theorem 1, and noting Eq. (\ref{factorizability condition}), to win the game, Alice and Bob joint strategy must yield outcomes which satisfy the following relation:
\begin{eqnarray}
&&\sum_{(\eta_{12},\xi)}\sum_{(\tilde{s}_{\vec{n}_1},\tilde{s}_{\vec{n}_2})}\sum_{(o_{\vec{n}_1},o_{\vec{n}_2})}~o_{\vec{n}_1}o_{\vec{n}_2}\nonumber\\
&\times&{\rm Pr}(o_{\vec{n}_1}|\tilde{s}_{\vec{n}_1},\eta_{12},\xi,\psi_{12}){\rm Pr}(o_{\vec{n}_2}|\tilde{s}_{\vec{n}_2},\eta_{12},\xi,\psi_{12})\nonumber\\
&\times&{\rm Pr}(\tilde{s}_{\vec{n}_1},\tilde{s}_{\vec{n}_2},\eta_{12},\xi|\vec{n}_1,\vec{n}_2,\psi_{12})\nonumber\\
&=&\braket{\psi_{12}|\hat{\sigma}_{\vec{n}_1}\otimes\hat{\sigma}_{\vec{n}_2}|\psi_{12}}.
\label{condition for winning pre}
\end{eqnarray}
Next, inserting Eq. (\ref{join factorizable distribution of continuum c-valued spins}) into Eq. (\ref{condition for winning pre}), we get, after summing over $\tilde{s}_{\vec{n}_{\mu}}$, $\mu=1,2$, 
\begin{eqnarray}
&&\sum_{(\eta_{12},\xi)}\sum_{(o_{\vec{n}_1},o_{\vec{n}_2})}~~o_{\vec{n}_1}o_{\vec{n}_2}\nonumber\\
&&\times{\rm Pr}(o_{\vec{n}_1}|\vec{n}_1,\eta_{12},\xi,\psi_{12}){\rm Pr}(o_{\vec{n}_2}|\vec{n}_2,\eta_{12},\xi,\psi_{12})\nonumber\\
&&\times{\rm Pr}(\eta_{12},\xi|\chi,\psi_{12})=\braket{\psi_{12}|\hat{\sigma}_{\vec{n}_1}\otimes\hat{\sigma}_{\vec{n}_2}|\psi_{12}}, 
\label{condition for winning}
\end{eqnarray}
where we have defined the conditional probabilities as
\begin{eqnarray}
&&{\rm Pr}(o_{\vec{n}_{\mu}}|\vec{n}_{\mu},\eta_{12},\xi,\psi_{12})\doteq\sum_{\tilde{s}_{\vec{n}_{\mu}}}{\rm Pr}(o_{\vec{n}_{\mu}}|\tilde{s}_{\vec{n}_{\mu}},\eta_{12},\xi,\psi_{12})\nonumber\\
&&\times\delta\Big(\tilde{s}_{\vec{n}_{\mu}};{\rm Re}\Big\{\frac{\braket{\eta_{12}|\hat{\sigma}_{\vec{n}_{\mu}}|\psi_{12}}}{\braket{\eta_{12}|\psi_{12}}}\Big\}+\frac{\xi}{\hbar}{\rm Im}\Big\{\frac{\braket{\eta_{12}|\hat{\sigma}_{\vec{n}_{\mu}}|\psi_{12}}}{\braket{\eta_{12}|\psi_{12}}}\Big\}\Big), \nonumber\\
\label{conditional local probability}
\end{eqnarray}
$\mu=1,2$. The condition for winning the game of Eq. (\ref{condition for winning}) therefore requires the two players to reconstruct the quantum spin correlation on the right-hand side, using a local hidden variable or local causal model given on the left-hand side. Noting this, when $\ket{\psi_{12}}$ is nonfactorizable, according to the Bell's theorem \cite{Bell's theorem,CHSH inequality}, no joint classical strategy of Alice and Bob are able to satisfy Eq. (\ref{condition for winning}). Namely, for the class of games wherein the initial correlation between $\tilde{s}_{\vec{n}_1}$ and $\tilde{s}_{\vec{n}_2}$ is equal to the quantum spins correlation over an entangled quantum state $\ket{\psi_{12}}$ (per Theorem 1), Alice and Bob outputs will always violate the constraint of conservation of correlation of Eq. (\ref{conservation of correlation}). We note that it needs an infinite number of rounds to be able to compute the correlation. One can however develop a winning criteria for a finite number of rounds, by looking at the convergence rate of the finite-ensemble correlation. 

As a concrete example, we can follow the CHSH set-up \cite{CHSH inequality} to run the game. Namely, at each round of the game, Charlie chooses one pair out of four alternative pairs of unit vectors, i.e., $(\vec{n}_1,\vec{n}_2)$, $(\vec{n}_1,\vec{n}'_2)$, $(\vec{n}'_1,\vec{n}_2)$ $(\vec{n}_1',\vec{n}'_2)$, randomly, and use them to compute a pair of c-values spin variables. Let us denote the correlation between the outputs, e.g., $o_{\vec{n}_1}$ and $o_{\vec{n}_2}$, i.e., Alice's output when she is given $\tilde{s}_{\vec{n}_1}$ and Bob's output when he is given $\tilde{s}_{\vec{n}_2}$, as $C_{\vec{n}_1\vec{n}_2}$. Then, assuming that the four pairs of unit vectors are sampled equally likely, if Alice and Bob joint strategy is classical, the correlation of their outputs must satisfy the Bell-CHSH inequality, i.e., $C_{12}^{\rm CHSH}\doteq C_{\vec{n}_1\vec{n}_2}+C_{\vec{n}_1\vec{n}'_2}+C_{\vec{n}'_1\vec{n}_2}-C_{\vec{n}'_1\vec{n}'_2}\le 2$. On the other hand, since the original correlation between real valued variables $\tilde{s}_{\vec{n}_1}$ and $\tilde{s}_{\vec{n}_2}$ is equal to the quantum spin correlation (as per Theorem 1), when $\ket{\psi_{12}}$ is entangled, computing the CHSH correlation $C_{12}^{\rm CHSH}$ for the associated c-valued spin variables will yield a value larger than 2 with a maximum value $2\sqrt{2}$. Hence, when the original correlation between the pair of real-valued variables correspond to a nonfactorizable vector $\ket{\psi_{12}}$, the fact that these correlations violate Bell inequalities says that no classical strategy of the joint mappings will ever win the game. 

If Alice and Bob have quantum circuits, and $\vec{n}_1,\vec{n}'_1,\vec{n}_2,\vec{n}'_2$ are coplanar so that their direction are parametered only by polar angles, then they can always win the game, by running the following strategy. First, they need to physically encode the complex vector $\ket{\psi_{12}}$ which generates the joint probability distribution of real-valued numbers of Eq. (\ref{join factorizable distribution of continuum c-valued spins}), as an ensemble of entangled pairs of spin-$\frac{1}{2}$ particles (or entangled pair of any physical qubits) in the quantum state $\ket{\psi_{12}}$. Alice then stores one of the particles in the entangled pairs to her circuit, and Bob the other particles in the pairs, and bring them to their separated labs. Next, Alice, upon receiving the triple $\{\eta_{12},\xi,\tilde{s}_{\vec{n}_1}\}$ from Charlie, infers $\vec{n}_1$ using the relation of  Eq. (\ref{real continuum c-valued spin variable}). For example, in the case when $\ket{\psi_{12}}$ is given by the singlet of Eq. (\ref{singlet state}) with the reference basis $\{\ket{\eta_{12}}\}=\{\ket{y+}\ket{x+},\ket{y+}\ket{x-},\ket{y-}\ket{x+},\ket{y-}\ket{x-}\}$, its polar angle $\theta_1$ can be easily inferred from $\{\eta_{12},\xi,\tilde{s}_{\vec{n}_1}\}$ using Eq. (\ref{c-valued spin for along theta for singlet}). Likewise, Bob, upon receiving the triple $\{\eta_{12},\xi,\tilde{s}_{\vec{n}_2}\}$ from Charlie, infers $\vec{n}_2$ using the relation of  Eq. (\ref{real continuum c-valued spin variable}). They then use the inferred unit vectors as the directions along which they make local spin measurements to their respective particles. Namely, Alice makes measurement of $\hat{\sigma}_{\vec{n}_1}$ to her particle, and similarly Bob makes measurement of $\hat{\sigma}_{\vec{n}_2}$ to his particle, yielding outcomes $\pm 1$ randomly. Alice assigns her outcomes to $o_{\vec{n}_1}$, and Bob to $o_{\vec{n}_2}$. In this sense, they map $(\tilde{s}_{\vec{n}_1},\tilde{s}_{\vec{n}_2})$ onto $(o_{\vec{n}_1},o_{\vec{n}_2})$, using the entangled particles and local spin measurement device. By construction, the correlation between $o_{\vec{n}_1}$ and $o_{\vec{n}_2}$ is given by the quantum spin correlation of $\braket{\psi_{12}|\hat{\sigma}_{\vec{n}_1}\otimes \hat{\sigma}_{\vec{n}_2}|\psi_{12}}$. Theorem 1 then guarantees that this correlation between the binary standard quantum spin values is equal to the original correlation between the continuum c-valued spins $\braket{\tilde{s}_{\vec{n}_1}(\eta_{12},\xi|\psi_{12})\tilde{s}_{\vec{n}_2}(\eta_{12},\xi|\psi_{12})}$. Hence, it satisfies the requirement to win the game, i.e., the constraint of conservation of correlation of Eq. (\ref{conservation of correlation}).     

It is thus clear that quantum entangled states are the nonclassical resource to win the above statistical game of spacelike joint mappings under correlation conservation. What is special about the mapping generated by the local spin measurement over the entangled quantum states so that it can be used to win the game while any classical strategy must fail? The basic assumption underlying the classical strategy is that the joint independent mapping of Eq. (\ref{coordinate mapping}) can be represented by the conditional probabilities ${\rm Pr}(o_{\vec{n}_{\mu}}|\tilde{s}_{\vec{n}_{\mu}},\eta_{12},\xi,\psi_{12})$, $\mu=1,2$ implying the factorizability condition of Eq. (\ref{factorizability condition}). Hence, the mapping generated by the local spin measurements over the entangled states somehow violates this plausible assumption, either by allowing nonlocal influence so that the conditional probability of $o_{\vec{n}_{\mu}}$ may depend on $o_{\vec{n}_{\nu}}$ or $\tilde{s}_{\vec{n}_{\nu}}$, $\nu\neq\mu$, or the mapping is acausal so that the above conditional probabilities simply cannot be defined. It is intriguing to pounder how the above game of joint mapping under conservation of correlation is related to what is really happening in the spin measurements in Bell-type experiment.

Finally, we emphasize that it is the requirement of conservation of correlation which forces any strategy to comply with the Bell theorem so that it must violate the Bell inequalities when $\ket{\psi_{12}}$ is nonfactorizable. We further note that while the protocol of the game of joint mapping is not as simple as the protocol of the CHSH game discussed at the beginning of this section, the requirement of conservation of correlation is much easier to grasp in physical terms than the winning condition of CHSH game of Eq. (\ref{CHSH condition}). Conservation of correlation appeals directly to intuition, and moreover, conservation laws have played prominent roles in the past in the construction of physical theories. 

\section{Conclusion}

The empirical violation of Bell inequalities \cite{Hensen loophole free Bell test,Giustina loophole free Bell test,Shalm loophole free Bell test} implies that we must give up, as measurement is concerned, at least one of the following: realism, locality, and free choice. This suggests that there must be a deep principle which measurement cannot resist obeying so that it is willing to sacrifice such intuitive and plausible concepts. To study this problem, we have assumed that a spin-$\frac{1}{2}$ particle (or any generic physical qubit) can always be assigned a definite c-valued spin variable regardless of measurement. The c-valued spin variable may take any continuum real value in the absence of measurement, and reduces to the binary values $\pm 1$ after the measurement reproducing the standard value of quantum spin. Moreover, the bipartite correlation of the c-valued spin variables prior to measurement is always equal to the quantum correlation obtained in quantum spin measurement. This motivates a speculation that quantum spin measurement maps the c-valued spin variables from continuous range of possible real number onto the binary $\pm 1$, while respecting the principle of conservation the correlation. In such a model, it is the plausible requirement of conservation of bipartite correlation which compels the measurement to violate the Bell inequalities when the prepared state is entangled.

We then constructed a statistical game of joint mappings which, to an extent, captures the above model of measurement. Alice and Bob, sufficiently faraway separated from each other, are asked to map, independently, a pair of real numbers sampled from a specific distribution, onto a pair of binary numbers $\pm 1$, with the condition that the statistical correlation is preserved. The winning condition of correlation conservation forces the game to comply with the Bell theorem which implies that, for certain class of the games associated with a nonfactorizable vector in Hilbert space, Alice and Bob can never win the game using any classical (i.e., local and deterministic) joint strategy. They can instead easily win the game with a quantum strategy using an ensemble of entangled spin-$\frac{1}{2}$ particles (qubits) and quantum circuits for local spin measurement. The game suggests that quantum protocols utilizing entanglement may exhibit quantum advantage | by way of violating Bell inequalities | in information processing tasks requiring conservation of bipartite correlation. 

\begin{acknowledgments}
This work is partly funded by Lembaga Penelitian dan Pengabdian Masyarakat, Institut Teknologi Bandung, under the program of research assignment with the contract number: 2971/IT1.B07.1/TA.00/2021. It is also in part supported by the Indonesia Ministry of Research, Technology, and Higher Education through PDUPT research scheme with the contract number: 2/E1/KP.PTNBH/2019. I would like to thank the anonymous Referees for constructive criticism and generous recommendation, and Hermawan K. Dipojono for useful discussion. 
\end{acknowledgments}

\end{document}